\def\isarxiv{1} 

\ifdefined\isarxiv
\documentclass[11pt]{article}

\usepackage[numbers]{natbib}

\else
\documentclass{article}
\usepackage{neurips_2022}
\fi

\usepackage{amsmath}
\usepackage{amsthm}
\usepackage{amssymb}
\usepackage{algorithm}
\usepackage{subfig}
\usepackage{algpseudocode}
\usepackage{graphicx}
\usepackage{grffile}
\usepackage{wrapfig,epsfig}
\usepackage{url}
\usepackage{xcolor}
\usepackage{epstopdf}

\usepackage{bbm}
\usepackage{dsfont}

\allowdisplaybreaks

\ifdefined\isarxiv

\usepackage{tikz}
\usepackage{hyperref}  
\hypersetup{colorlinks=true,citecolor=blue,linkcolor=blue} 
\usetikzlibrary{arrows}
\usepackage[margin=1in]{geometry}

\else

\usepackage{microtype}
\usepackage{hyperref}
\definecolor{mydarkblue}{rgb}{0,0.08,0.45}
\hypersetup{colorlinks=true, citecolor=mydarkblue,linkcolor=mydarkblue}

\fi

\newtheorem{theorem}{Theorem}[section]
\newtheorem{lemma}[theorem]{Lemma}
\newtheorem{definition}[theorem]{Definition}

\newcommand{\wh}{\widehat}

\newcommand{\ov}{\overline}
\newcommand{\N}{\mathcal{N}}
\newcommand{\R}{\mathbb{R}}

\renewcommand{\hat}{\wh}

\DeclareMathOperator*{\Z}{\mathbb{Z}}

\DeclareMathOperator{\OPT}{OPT}

\DeclareMathOperator{\poly}{poly}

\DeclareMathOperator{\nnz}{nnz}

\DeclareMathOperator{\rank}{rank}

\DeclareMathOperator{\vect}{vec}

\makeatletter
\newcommand*{\RN}[1]{\expandafter\@slowromancap\romannumeral #1@}
\makeatother

\usepackage{lineno}

\begin{document}

\ifdefined\isarxiv

\date{}

\title{Solving Tensor Low Cycle Rank Approximation}
\author{
Yichuan Deng\thanks{\texttt{ycdeng@cs.washington.edu}. The University of Washington.}
\and
Yeqi Gao\thanks{\texttt{a916755226@gmail.com}. The University of Washington.}
\and
Zhao Song\thanks{\texttt{zsong@adobe.com}. Adobe Research.}
}

\else

\title{Intern Project} 
\maketitle 
\fi

\ifdefined\isarxiv
\begin{titlepage}
  \maketitle
  \begin{abstract}

Large language models have become ubiquitous in modern life, finding applications in various domains such as natural language processing, language translation, and speech recognition. Recently, a breakthrough work [Zhao, Panigrahi, Ge, and Arora Arxiv 2023] explains the attention model from probabilistic context-free grammar (PCFG). 
One of the central computation task for computing probability in PCFG is formulating a particular tensor low rank approximation problem, we can call it tensor cycle rank. Given an $n \times n \times n$ third order tensor $A$, we say that $A$ has cycle rank-$k$ if there exists three $n \times k^2$ size matrices $U , V$, and $W$ such that for each entry in each 
\begin{align*}
A_{a,b,c} = \sum_{i=1}^k \sum_{j=1}^k \sum_{l=1}^k U_{a,i+k(j-1)} \otimes V_{b, j + k(l-1)} \otimes W_{c, l + k(i-1) }
\end{align*}
for all $a \in [n], b \in [n], c \in [n]$.
For the tensor classical rank, tucker rank and train rank, it has been well studied in [Song, Woodruff, Zhong SODA 2019]. In this paper, we generalize the previous ``rotation and sketch'' technique in page 186 of [Song, Woodruff, Zhong SODA 2019] and show an input sparsity time algorithm for cycle rank.

  \end{abstract}
  \thispagestyle{empty}
\end{titlepage}

{\hypersetup{linkcolor=black}
}
\newpage

\else

\begin{abstract}

Large language models have become ubiquitous in modern life, finding applications in various domains such as natural language processing, language translation, and speech recognition. Recently, a breakthrough work [Zhao, Panigrahi, Ge, and Arora Arxiv 2023] explains the attention model from probabilistic context-free grammar (PCFG). 
One of the central computation task for computing probability in PCFG is formulating a particular tensor low rank approximation problem, we can call it tensor cycle rank. Given an $n \times n \times n$ third order tensor $A$, we say that $A$ has cycle rank-$k$ if there exists three $n \times k^2$ size matrices $U , V$, and $W$ such that for each entry in each 
\begin{align*}
A_{a,b,c} = \sum_{i=1}^k \sum_{j=1}^k \sum_{l=1}^k U_{a,i+k(j-1)} \otimes V_{b, j + k(l-1)} \otimes W_{c, l + k(i-1) }
\end{align*}
for all $a \in [n], b \in [n], c \in [n]$.
For the tensor classical rank, tucker rank and train rank, it has been well studied in [Song, Woodruff, Zhong SODA 2019]. In this paper, we generalize the previous ``rotation and sketch'' technique in page 186 of [Song, Woodruff, Zhong SODA 2019] and show an input sparsity time algorithm for cycle rank.

\end{abstract}

\fi

\section{Introduction}

Large Language Models (LLMs) are a type of artificial intelligence model 
that is designed to process and understand human language. These models are built using deep learning techniques and can be trained on massive amounts of data, allowing them to learn patterns and relationships in language that enable them to generate text, answer questions and complete other tasks that are related to language. Some of the most well-known LLMs include OPT \cite{zrg+22}, PaLM \cite{cnd+22},  GPT-3 \cite{bmr+20}, Transformer \cite{vsp+17} and BERT \cite{dclt18}, which have been used for a wide range of applications such as language translation \cite{s20,fbs+21,lgg+20}, and content generation \cite{vsp+17,dclt18,bmr+20,cnd+22}. Large language models are becoming increasingly sophisticated and powerful, and are expected to play a major role in shaping the future of human-computer interaction and natural language processing \cite{wds+19,rpsw19,qsx+20,lyf+23}.

As one of the large language models, the Transformer \cite{vsp+17} has drawn the most attention in recent years. Recently, a beautiful and breakthrough work by Zhao, Panigrahi, Ge and Arora \cite{zpga23} explore an interesting problem regarding whether the Transformer is capable of true parsing while attempting masked word prediction. 

For solving the problem above, they first focus a Probabilistic Context-Free Grammar (PCFG) \cite{y60} and parsing algorithm based on it.
As a type of formal grammar, Probabilistic Context-Free Grammar (PCFG) is widely used in natural language processing \cite{wds+19,qsx+20} and computational linguistics \cite{m22,bds19}. 
With $\mathcal{N}$ as the set of all the non-terminal symbols of the language grammar, $\mathcal{P}\subset \mathcal{N}$ as the set of pre-terminals, $\mathcal{I} \subset \mathcal{N}$ as the finite set of in-terminals and $[n]$ as the set of all possible words and $p$ is the relating probability for grammar parsing, a $5$-tuple $\mathcal{G} := (\mathcal{N},\mathcal{P},\mathcal{I},p,n)$ defines it. 

In a PCFG, every grammar rule of the language is defined to be connected with a probability that represents how likely it is to be used in generating a sentence. These probabilities can be learned from a large corpus of text data using machine learning techniques such as the expectation-maximization algorithm. PCFGs have been widely used in various applications of natural language processing, including part-of-speech tagging \cite{c89,tkm03,r96}, syntactic parsing  \cite{c96,km03,pk07}, and text generation \cite{lk98}. 

A well-known approach for discovering the (unlabeled) parse tree of a sentence $w$ using the PCFG model is the Labelled-Recall algorithm, which was introduced by \cite{g96}. 
The algorithm aims to identify the tree $T = {(i, j)}$ that maximizes the sum of scores, where the score of a span $v_i v_{i+1} \cdots v_j$ under a non-terminal $a$ is defined as the marginal probability $\mu(i, j,a)$ calculated as follows: $\mu(i, j,a) = \max_{a \in \mathcal{N}} \Pr[a \rightarrow v_i v_{i+1}\cdots v_j, \mathsf{root} \rightarrow w | \mathcal{G}]$. 

The marginal probabilities mentioned above can be computed using the Inside-Outside algorithm \cite{ms99,b79}. In a recent study, \cite{zpga23} attempted to explain the Transformer architecture from the perspective of Probabilistic Context-Free Grammar (PCFG) \cite{msm93}.

With the aid of dynamic programming, the Inside-Outside algorithm can determine two types of probability terms. The inside type of probability is defined as 
\begin{align*}
    p_{\mathrm{in}}(i,j,a) := \Pr[a\to v_iv_{i+1}\dots v_j|\mathcal{G}]
\end{align*}
while the outside type of probability is defined as 
\begin{align*}
    p_{\mathrm{out}}(i, j,a)=\Pr[\mathsf{root}\rightarrow v_1v_2\dots v_{i-1} a v_{j+1}\dots v_L|\mathcal{G}]
\end{align*}
for all ordered pairs $(i, j)$ where $1 \le i \le j \le L$ and $a \in \mathcal{N}$. To be precise, the probabilities can be defined with the following recursive form,
{\small
\begin{align}\label{eq:inside_probability}
    p_{\mathrm{in}}(i,j,a)  = \sum_{b,c}\sum_{k=i}^{j-1} p_{\mathrm{in}}(i,k,b)p_{\mathrm{in}}(k+1,j,c)\Pr[a\to bc] \notag
\end{align}
}
and 
\begin{align}
     p_{\mathrm{out}}(i,j,a)  
    =  & ~ \sum_{b, c} \sum_{k=1}^{i-1}p_{\mathrm{in}}(k, i-1,c) p_{\mathrm{out}}(k, j,b) \Pr[b \to c a] \notag \\
    + & ~ \sum_{b, c} \sum_{k=j+1}^{L}p_{\mathrm{in}}(j+1, k,c) p_{\mathrm{out}}(i, k,b)\Pr[b \to a c]  \notag
\end{align}
given that for all $a,i$, $p_{\mathrm{in}}(i,i,a) = \Pr[a\to v_i]$ and  for all $a$, $p_{\mathrm{out}}(1,L,\mathsf{root})=1$. We can obtain the marginal probabilities using the following expressions
\begin{align}
    \mu(i,j,a) 
    = & ~ \Pr[a\Rightarrow v_iv_{i+1}\dots v_j, \mathsf{root}\Rightarrow w|\mathcal{G}] \notag \\
    = & ~ \Pr[\mathsf{root}\rightarrow v_1v_2\dots v_{i-1} a v_{j+1}\dots v_L|\mathcal{G}] \cdot \Pr[a\to v_iv_{i+1}\dots v_j|\mathcal{G}] \notag \\
    = & ~ p_{\mathrm{in}}(i,j,a)\cdot p_{\mathrm{out}}(i,j,a). \notag
\end{align}

In our paper, we will focus on the analysis of $p_{\mathrm{in}}(i,j,a)$ and investigate  $p_{\mathrm{in}}(i,j,b)$, $p_{\mathrm{in}}(k+1,j,c)$ and $\Pr[a\rightarrow bc]$ from a tensor perspective. By treating $b,c,k$ as the cycle index and $a,i,j$ as the tensor index transformed from $p_{\mathrm{in}}(i,j,a)$, we can restructure the function in a new format. Drawing inspiration from the computation of $p_{\mathrm{in}}(i,j,a)$, we define the cycle rank of a tensor for low-rank approximation as follows. 
\begin{definition}[Tensor Cycle-rank]
 Considering $\ov{U}$, $\ov{V}$, and $\ov{W}$, which are $n$-by-$k$-by-$k$ tensors with real-valued entries, are the tensorizations  
 of matrices $U\in \R^{n \times k^2}$, $V \in \R^{n \times k^2}$, and $W \in \R^{n \times k^2}$, we say $a$ has cycle-rank provided that matrices $U, V,$ and $W$ belong to $\R^{n \times k^2}$ satisfying the following condition:
\begin{align*}
A_{i,j,l} = \sum_{a=1}^k \sum_{b=1}^k \sum_{c=1}^k U_{i, a + k(b-1)} V_{j, b + k(c-1)} v_{l, c + k(a-1)}
\end{align*}
Alternatively, we can state that
\begin{align*}
A_{i,j,l} = \sum_{a=1}^k \sum_{b=1}^k \sum_{c=1}^k \ov{U}_{i, a , b} \ov{V}_{j, b ,c } \ov{W}_{l, c , a}
\end{align*}
\end{definition}

We remark that there are other tensor ranks such as classical tensor rank, tensor train rank, and tensor tucker rank. Those ranks have been studied by \cite{swz19}, however tensor cycle rank is not studied. We provide more detailed definition of other tensor ranks in Section~\ref{sec:preli:tensor_ranks}.

\subsection{Our Result}
We present our main result as follows:
\begin{theorem}[Cycle rank, informal version of Theorem~\ref{thm:main_formal}]\label{thm:main_informal}
 For any $k\geq 1$, $\epsilon \in (0,1)$, and a third-order tensor $a\in \R^{n\times n\times n}$, there is an algorithm that operates in 
\begin{align*}
O(\nnz(A)) + n \poly(k,1/\epsilon) + \exp( O( \poly(k)/\epsilon ))
\end{align*}
time and produces $U$, $V$, and $W$ which are tensors of dimension $n \times k \times k$ satisfying the following condition:
\begin{align*}
\| \sum_{i=1}^k \sum_{j=1}^k \sum_{l=1}^k U_{*,i,j} \otimes V_{*,j,l} \otimes v_{*,l,i} - A  \|_F \leq (1+\epsilon) \underset{\mathrm{cycle}~\rank-k~ A_k }{\min} \| A_k - A\|_F^2
\end{align*}

\end{theorem}

\paragraph{Roadmap.}
We divide the paper as follows. In Section~\ref{sec:rel_work} we discuss some related work. In Section~\ref{sec:tech_ov} we give the overview of our techniques. In Section~\ref{sec:prel} we provide preliminaries for our work. In Section~\ref{sec:tensor_rank} we give our formal result and analysis.

\section{Related Work}
\label{sec:rel_work}

\subsection{Transformer Theory}

\paragraph{Optimization and Convergence}

In the realm of optimization, \cite{szks21} concentrated on investigating the behavior of a single-head attention mechanism to emulate the process of learning a Seq2Seq model, while adaptive methods have been emphasized for attention models by \cite{zkv+20}. \cite{gms23} have recently become interested in the over-parametrization problem associated with exponential activation functions.
The exponential activation function is the focus of study in \cite{gms23}, where the convergence of an over-parameterized two-layer neural network is examined. The authors determine the minimum number of neurons required for convergence.
Furthermore, \cite{lsz23} have addressed the issue of regularized exponential regression and proposed an algorithm running in input sparsity time. \cite{llr23} provides a comprehensive explanation of how transformers learn ``semantic structure", which is in regard to their proficiency in capturing word co-occurrence patterns.

\paragraph{Fast Attention Computation}

The work of \cite{zhdk23} and \cite{as23} primarily focus on the static version of attention computation. In particular, on the positive side, \cite{as23} provides an algorithm, and on the negative side, \cite{as23} prove a hardness by assuming (strong) exponential time hypothesis.  \cite{bsz23} define and investigate the dynamic version of the problem, presenting both positive and negative results. The algorithmic result of \cite{bsz23} is built on lazy update techniques in solving linear programming \cite{cls19,b20}. The hardness result of \cite{bsz23} is built on Hinted MV conjecture \cite{bns19}. \cite{dms23} consider the case where feature dimension is much larger than the length of sentence in attention computation, and they provide both randomized algorithm and deterministic algorithm to sparsify the feature dimension.

\paragraph{Learning in-context}
In a recent work \cite{asa+22}, it is shown that, in-context learners based on transformers can perform conventional learning algorithms in an implicit manner by encoding smaller models within their activations and continuously updating these models as new examples are presented within the context. 
Concentrating on the clearly defined task of training a model under in-context conditions to learn a class of functions, such as linear functions, \cite{gtlv22} aim to gain a deeper insight into in-context learning. Their research aims to determine if a model, when provided with data derived from some functions within a class, are able to be trained to learn the "majority" of functions in this class. \cite{onr+22} gave an explanation of the mechanisms of Transformers as in-context learners. They showed that the training process of the Transformers in in-context tasks have similarity of some meta-learning formulations based on gradient descent.

\paragraph{Privacy.}
Considering the potential negative influence of the abuse of the LLMs, 
\cite{kgw+23} propose a method for watermarking proprietary language models without adversely affecting text quality. A novel algorithm for detecting watermarks in language models without requiring access to the model's parameters or API is introduced by \cite{kgw+23}.

\subsection{Numerical Linear Algebra}

\paragraph{Low-rank Approximation}

Many variants of low-rank approximation problems have been studied, for example low-rank approximation with Frobenious norm \cite{cw13,nn13}, matrix CUR decomposition \cite{bw14,swz17,swz19}, weighted low rank approximation \cite{rsw16}, entrywise $\ell_1$ norm low-rank approximation \cite{swz17,swz19_neurips_l1}, general norm column subset selection \cite{swz19_neurips_general}, tensor low-rank approximation \cite{swz19}, and tensor regression \cite{dssw18,djs+19,rsz22,swyz21}. Low-rank approximation also has many applications in other problems such as cutting plane method \cite{jlsw20}, integral minimization \cite{jlsz23}, training neural network \cite{bpsw21,szz21,z22}.

\paragraph{Input Sparsity Algorithms}
For several years, there have been many works focused on designing input-sparsity time algorithms \cite{cw13, nn13, wz16, bwz16, rsw16, swz17, swz19, syyz22, dsw22}. For the problems of matrix low-rank approximation and linear regression, \cite{cw13} gave the first input sparsity time. With a sparse embedding matrix $S$, one can compute its product with any input matrix $A$ in the input-sparsity $\nnz(A)$ time. 
Another work \cite{nn13} later gave some sketching matrices called Oblivious Sparse Norm Approximating Projections (OSNAPs), which improve the former dimension of sketch in \cite{cw13}. \cite{rsw16} gives an input sparsity time algorithm for weighted low-rank approximation. \cite{swz17} generalize the Frobenius low rank approximation problem \cite{cw13,nn13} to entry-wise $\ell_1$ low rank approximation problem. \cite{swz19} studied the tensor low rank approximation problem. Recently, \cite{syyz22} gave an algorithm for structured John Ellipsoid computation. It improves the previous work \cite{ccly19} to input sparsity time. Another recent work \cite{dsw22} gave a novel algorithm which can minimizing the discrepancy in input sparsity time. 

\section{Technique Overview}
\label{sec:tech_ov}
With  $U^*\in \R^{n^2 \times k^2}$, $V^* \in \R^{n^2 \times k^2}$ and $W^* \in \R^{n^2 \times k^2}$, we can rewrite $A_k$ as follows 

\begin{align*}
A_k = \sum_{i=1}^k \sum_{j=1}^k \sum_{l=1}^k U_{i+k (j-1)}^* \otimes V_{j+k(l-1)}^* \otimes W_{l+k(i-1)}^*
\end{align*}

The main idea in our proof is to first establish the existence of a matrix decomposition for $A$, and then use a corresponding algorithm to confidently locate the desired matrix.
\paragraph{Rotate and Sketch Step 1.}
In the first step, we consider the hypothetical regression problem $\min_{U\in \R^{n\times k^2}} \|A_1 - UZ_1\|_F^2$. With $r = O(k^2/\epsilon)$ and $S_1\in \R^{n^2 \times r}$  with i.i.d. normal random variables denoted $N(0, 1/r)$, and $\hat{U} = \textrm{argmin}_{U\in \R^{n\times k^2}}\|UZ_1S_1 - A_1 S_1\|_F^2$, we have
\begin{align*}
    \|A_1 -\hat{U}Z_1\|_F^2 \leq (1+\epsilon) \min_{U\in \R^{n \times k^2}} \|A_1 - UZ_1\|_F^2. 
\end{align*}
By choosing $\hat{U} = A_1 S_1 (Z_1 S_1)^{\dagger}$, we can prove that at least a matrix $\hat{U}$ can be found to satisfy the equation above.

\paragraph{Rotate and Sketch Step 2.}
According to the first step, we can create a rank-$k$ tensor, defined as follows 
\begin{align*}
B := \sum_{i=1}^k \sum_{j=1}^k \sum_{l=1}^k \hat{U}_{i+k(j-1)} \otimes V^*_{j+k(l-1)} \otimes W^*_{j+k(i-1)}
\end{align*}
from the matrix $\hat{U}Z_1$. 

Since $\hat{U}$ is in the column span of $A_1 S_1$, the set of rank-$1$ tensors $\{( A_1 S_1 )_{a+k(b-1)} \otimes V^*_{b+k(c-1)} \otimes W^*_{c+k(a-1)}\}_{a \in [r], b,c \in [k]}$ spans the space that contains $B$. It is possible to express a proper rank-$k$ tensor $B$ as a linear combination of rank-$1$ tensors of the form $\{U^*_{a+k(b-1))} \otimes (A_2 S_2)_{b+k(c-1)} \otimes W^*_{c+k(a-1)}\}_{a, c \in [k], b \in [r]}$ and the set of rank-$1$ tensors $\{U^*_{a+k(b-1)} \otimes V^*_{b+k(c-1)} \otimes (A_3 S_3 )_{c+k(a-1)}\}_{a,b \in [k], c \in [r]}$ spans a subspace that contains a proper rank-$k$ tensor $B$.

We first compute $A_1 S_1$, and write $\hat{U} = A_1 S_1 (Z_1 S_1)^{\dagger}$. Now, we redefine $Z_2$ {\it with respect to $\hat{U}$}, which implies that, the rows of $Z_2$ is vectors where the $(j,l)$-th row is $\vect(\sum_{i=1}^k \wh{U}_{i+k(j-1)} \otimes W^*_{l+k(i-1)})$ for each $j \in [k]$ and $l \in [k]$. And we consider the regression problem in the following form,
\begin{align*}
    \min_V \|VZ_2-A_2\|_F^2.
\end{align*}
If $S_2 \in \R^{n^2 \times r}$ as a matrix of i.i.d. Gaussian entries, we have $\hat{V} = A_2 S_2 (Z_2 S_2)^{\dagger}$ satisfies 
\begin{align*}
    \|\hat{V} Z_2-A_2\|_F^2 \leq (1+\epsilon)^2 \|A_k-A\|_F^2. 
\end{align*}
The matrix $B$ can be obtained by retensorizing $\hat{V}Z_2$ and then we will have
\begin{align*}
    \|B-A\|_F^2 = \|\hat{V} Z_2-A_2\|_F^2 \leq (1+\epsilon)^2 \|A_k-A\|_F^2.
\end{align*}

\paragraph{Rotate and Sketch Step 3.}
Recall that, the columns of $\hat{V}$ are all in the row span of $A_2 S_2$, and the rows of $Z_2$ are defined to be $\vect(\sum_{i=1}^k \wh{U}_{i+k(j-1)} \otimes W^*_{l+k(i-1)})$ for each $j \in [k]$ and $l \in [k]$, where the columns
 of $\hat{U}$ are in the row span of the matrix $A_1 S_1$, we have that, $B$ is in the span of the set of rank-$1$ tensors
\begin{align*}
  \{(A_1 S_1)_{a+k(b-1)} \otimes ( A_2 S_2 )_{b+k(c-1)} \otimes \hat{V}_{c+k(a-1)}\}_{a,b \in [r], c \in [k]}.
\end{align*}

Given that $Z_3 \in \R^{r^2 \times n^2}$ as an  matrix with rows $\vect( ( A_1 S_1)_{a+k(b-1)} \otimes ( A_2 S_2)_{b+k(c-1)} )$ for all pairs $a,b \in [r]$, we will focus on the following regression problem 
\begin{align*}
    \min_W \|WZ_3-A_3\|_F^2.
\end{align*}
By solving another regression problem, $W \in \R^{n \times r^2}$ can be solved in polynomial time. After retensorizing $WZ_3$ to a new tensor $B$, we can find a rank-$r^2 = O(k^2/\epsilon^2)$ tensor $B$ such that
\begin{align*}
\|B-A\|_F^2 \leq (1+\epsilon)^2 \|A_k-A\|_F^2 = (1+O(\epsilon))\|A_k-A\|_F^2.     
\end{align*}  

\paragraph{Input sparsity reduction.}
To obtain the input-sparsity running time guarantee of $n \poly(k/\epsilon)+\nnz(A)$ for Theorem~\ref{thm:f_main_algorithm}, while we can replace $S_1$ and $S_2$ using a sparse CountSketch matrix together with a Gaussian matrix, which enables us to compute $A_1 S_1$ and $A_2 S_2$ in the desired $\nnz(A) + n \poly(k/\epsilon)$ time.
  
For the next steps of solving $\min_W \|A_3 - WZ_3\|_F^2$ quickly and writing matrix $Z_3$ without spending $r^2 n^2$ time, a technique of TensorSketch is used, which is introduced in \cite{p13,pp13}. But as shown in \cite{anw14}, we need show some stronger properties of approximate matrix product ({\sf AMP}) and subspace embedding ({\sf SE}) to make it stand. When the AMP and SE is satisfied, we can solve the ``sketched version" regression problem, in the form of
\begin{align*}
    \min_W \|A_3S_3 - WZ_3S_3\|_F^2.
\end{align*}
The point of this sketch is, matrix products $A_3S_3$ and $Z_3S_3$ can be computed in the sparsity time of $\nnz(A) + n \poly(k/\epsilon)$. Combining them together, we can solve the problem in $n \poly(k/\epsilon)$ time. 

In order to get the sparsity time of $A$, we use CounSketch matrices to boost the algorithm. To be specific, we choose three CountSketch matrices $T_1, T_2, T_3$ with sketching dimension of $t = n \cdot \poly(k,1/\epsilon)$. 

We define  
\begin{align*}
D:=\sum_{i=1}^k\sum_{j=1}^k\sum_{l=1}^k (T_1 A_1 S_1 X_1)_{a,i+k(j-1)} \cdot (T_2 A_2 S_2 X_2)_{b,j+k(l-1)} \cdot (T_3 A_3 S_3 X_3)_{c,l+k(i-1)}.
\end{align*}
Applying them to the iterative argument described above, now we try to minimize the following target polynomial function
\begin{align*}
    \sum_{a,b,c \in [t]} ( D- (A(T_1,T_2,T_3))_{a,b,c})^2
\end{align*}
by choosing $3rk = O(k^2/\epsilon)$ indeterminates $X_1, X_2, X_3$. 

\paragraph{Solving small problem via polynomial system solver.}
  With $X_1, X_2$, and $X_3$ as $r \times k^2$ matrices of indeterminates, we can re-express $\hat{U}$ as $A_1 S_1 X_1$, $\hat{V}$ as $A_2 S_2 X_2$, and $\hat{W}$ as $A_3 S_3 X_3$. We let the rows of $Z_1$ be $\vect(V^*_{j+k(l-1)} \otimes W^*_{l+k(i-1)})$, the rows of $Z_2$ be $\vect(\hat{U}_{i+k(j-1)} \otimes W^*_{j+k(l-1)})$, and the rows of $Z_3$
  be  $\vect(\hat{U}_{i+k(j-1)} \otimes V_{j+k(l-1)})$. 
  
  It follows that the rank-$k$ tensor  
  \begin{align*}
  B =\sum_{i=1}^k \sum_{j=1}^k \sum_{l=1}^k (A_1 S_1 X_1)_{i + k (j-1) } \otimes (A_2S_2 X_2)_{j+k(l-1)} \otimes ( A_3S_3 X_3 )_{l+ k (i-1)}        
  \end{align*}
 where $\|A-B\|_F^2 \leq (1+\epsilon)^3 \|A-A_k\|_F^2$. 

The minimization task can be achieved by minimizing the degree-$6$ polynomial as follows
    \begin{align*}
        \sum_{a,b,c \in [n]} (\sum_{i=1}^k\sum_{j=1}^k\sum_{l=1}^k (A_1 S_1 X_1)_{a,i+k(j-1)} \cdot (A_2 S_2 X_2)_{b,j+k(l-1)} \cdot (A_3 S_3 X_3)_{c,l+k(i-1)} - A_{a,b,c}  )^2
    \end{align*}
by choosing $3rk = O(k^2/\epsilon)$ indeterminates $X_1, X_2, X_3$. We can get $X_1,X_2,X_3$ such that the solution is at least $(1+O(\epsilon))$-approximate.
 
Assuming that the entries of matrices $U^*, V^*, W^*$ are all bounded by $2^{\poly(n)}$, as stated in Theorem~\ref{thm:f_solving_small_problems}, the polynomial can be minimized up to an additive error of $2^{-\poly(n)}$ in $\poly(n)$ time \cite{r92,bpr96}. Similar approaches can also be used to obtain a relative error approximation of the ground truth value of $\OPT$, and to deal with situations where $A_k$ does not exist.

\section{Preliminary}
\label{sec:prel}
In this section we give preliminaries for our work. In Section~\ref{sec:notation}, we introduce our notations. In Section~\ref{sec:preli:tensor_ranks}, we provide definitions for different kinds of tensor ranks. In Section~\ref{sec:cycle_def}, we introduce definitions about the tensor cycle ranks. In Section~\ref{sec:pre_subspace_embeddings_approximate_matrix_product}, we introduce definitions of subspace embedding and approximate matrix product. In Section~\ref{sec:gauss_countsketch} we introduce results about the sketching techniques. In Section~\ref{sec:f_1_plus_epsilon}, we introduce the tensor low-rank approximation algorithm. In Section~\ref{sec:f_input_sparsity_reduction}, we introduce the technique for input-sparsity approximation. In Section~\ref{sec:f_solving_small_problems} we introduce a previous result for solving small problems. 

\subsection{Notations}
\label{sec:notation}

We use the notation $[n]$ to denote the set ${1,2,\cdots, n}$ for any positive integer $n$. Additionally, for a vector $x \in \R^n$, $\| x \|_2$ represents its $\ell_2$ norm, which is defined as $\| x \|_2: = (\sum{i=1}^n x_i^2 )^{1/2}$.
For a matrix $A \in \R^{n \times n}$, we use $\| A \|_F$ to denote its Frobenius norm, i.e., $\| A \|_F:= ( \sum_{i=1}^n \sum_{j=1}^n A_{i,j}^2 )^{1/2}$. For a tensor $A$, we use $\| A \|_F$ to denote its Frobenius norm, i.e., $\| A \|_F: = ( \sum_{i=1}^n \sum_{j=1}^n \sum_{l=1}^n A_{i,j,l}^2 )^{1/2}$.

For any tensor $A$, we use $\nnz(A)$ to denote the number of non-zero entries in tensor $A$.

For an invertible and square matrix $A$, we use $A^{-1}$ to denote its true inverse.

Let the SVD of $A \in \R^{n \times k}$ to be $M \Sigma N^\top$ where $M \in \R^{n \times k}$ and $N \in \R^{k \times k}$ have orthonormal columns, and $\Sigma \in \R^{k \times k}$ is a diagonal matrix. We denote the Moore-Penrose pseudoinverse matrix $A$ as $A^{\dagger} \in \R^{k \times n}$, i.e., $A^{\dagger} = N \Sigma^{-1} M^\top$.

We define the following two definitions of the $\otimes$ product. 
\begin{definition}[$\otimes$ product for vectors]
Given three vectors $a, b, c \in \R^n$, $a \otimes b \otimes c$ is used to denote a size $n \times n \times n$ tensor where $i,j,l$-th entry is $a_i b_j c_l$.
\end{definition}

\begin{definition}[$\otimes$ product for matrices]
\label{def:otimes_product}
    For $m$ matrices $A_1\in \R^{n_1 \times k}$, $A_2 \in \R^{n_2 \times k}$, $\cdots$, $A_m\in \R^{n_m\times k}$, we denote the tensor product in the following form
    \begin{align*}
        A_1 \otimes A_2 \otimes \cdots \otimes A_q = \sum_{i=1}^k (A_1)_i \otimes (A_2)_i \otimes \cdots \otimes (A_q)_i \in \R^{n_1 \times n_2 \times \cdots \times n_m},
    \end{align*}
    where we use $(A_j)_i$ to denote the $i$-th column of the matrix $A_j$. 
\end{definition}
We define the following $\vect()$ operation of converting a tensor into a vector, 
\begin{definition}[$\vect()$, converting a tensor to a vector]\label{def:vect}
    For any tensor $T \in \R^{n_1 \times n_2 \times \cdots \times n_q}$, we define
    \begin{align*}
        \vect(T)\in\R^{1\times \prod_{i=1}^q n_i}
    \end{align*}
    to be a row vector. For $t = (j_1-1)\prod_{i=2}^q n_i+(j_2-1)\prod_{i=3}^q n_i+\cdots+(j_{q-1}-1) n_q+j_{q}$, where $j_1 \in [n_1], j_2 \in [n_2] \dots, j_q \in [n_q]$, the $t$-th entry of $\vect(T)$ is $T_{j_1,j_2,\cdots,j_q}$.
\end{definition}

As an illustration, given that $u=\begin{bmatrix}5\\4\end{bmatrix},v=\begin{bmatrix}3\\2\\1\end{bmatrix}$ then $\vect(u\otimes v)=\begin{bmatrix}15&10&5&12&8&4\end{bmatrix}.$

\subsection{Tensor Ranks}
\label{sec:preli:tensor_ranks}
In this section, we define the following kinds of ranks for tensor. 
\begin{definition}[Classical rank]
We say tensor $U \in \R^{n \times n \times n}$ has a classical rank of $k$ if $k \in \Z_{\ge 0}$ is the smallest positive integer for which there exist three matrices $A, B, C \in \R^{n \times k}$ such that
\begin{align*}
    U_{i,j,l} = \sum_{a=1}^k \sum_{b=1}^k \sum_{c=1}^k A_{i, a} B_{j, b} C_{l, c}
\end{align*}
\end{definition}
We define Tucker rank as follows:
\begin{definition}[Tucker rank]
We say tensor $U \in \R^{n \times n \times n}$ has a tucker rank of $k$ if $k \in \Z_{\ge 0}$ is the smallest positive integer such that there exist three matrices $A\in \R^{n\times k}, B\in \R^{n\times k},C \in \R^{n\times k}$, and a (smaller) tensor $D \in \R^{k\times k \times k}$  such that $\forall (i,j,l) \in [n]\times[n]\times[n]$, we have 
\begin{align*}
    U_{i,j,l} = \sum_{i'=1}^k \sum_{j'=1}^k \sum_{l'=1}^k D_{i',j',l'} A_{i,i'} B_{j,j'} C_{l,l'}.
\end{align*}
\end{definition}

Then, we define train rank,
\begin{definition} [Train rank]
We say tensor $U \in \R^{n \times n \times n}$ has a train rank of $k$ if $k \in \Z_{\ge 0}$ is the smallest positive integer such that there exist three matrices $A \in \R^{1 \times n \times k}$, $B\in \R^{k\times n \times k}$, $C\in \R^{k\times n\times 1}$ such that $\forall i,j,l \in [n] \times [n] \times [n]$ we have,
\begin{align*}
     U_{i,j,l} = \sum_{i_1=1}^1 \sum_{i_2=1}^k \sum_{i_3=1}^k \sum_{i_4=1}^1 A_{i_1,i,i_2} B_{i_2,j,i_3} C_{i_3,l,i_4}
\end{align*}
\end{definition}

\subsection{Tensor Cycle Definitions}
\label{sec:cycle_def}
We give the formal definition of tensor cycle rank as follows. 

\begin{definition}[Cycle-rank]\label{def:cycle_rank}
We say $U \in \R^{n \times n \times n}$ has cycle-rank $k$ if there are three matrices $A, B, C \in \R^{n \times k^2}$ such that
\begin{align*}
U_{i,j,l} = \sum_{a=1}^k \sum_{b=1}^k \sum_{c=1}^k A_{i, a + k(b-1)} B_{j, b + k(c-1)} C_{l, c + k(a-1)}
\end{align*}
or
\begin{align*}
A_{i,j,l} = \sum_{a=1}^k \sum_{b=1}^k \sum_{c=1}^k \ov{U}_{i, a , b} \ov{V}_{j, b ,c } \ov{W}_{l, c , a}
\end{align*}
where $\ov{U} \in \R^{n \times k \times k}$ is tensorization of $U \in \R^{n \times k^2}$,  $\ov{V} \in \R^{n \times k \times k}$ is tensorization of $V \in \R^{n \times k^2}$, $\ov{W} \in \R^{n \times k \times k}$ is tensorization of $W \in \R^{n \times k^2}$.
\end{definition}

\subsection{Approximate Matrix Product and Subspace Embeddings}
\label{sec:pre_subspace_embeddings_approximate_matrix_product}

We define subspace embedding \cite{s06} as follows.

\begin{definition}[Subspace Embedding \cite{s06}]
\label{def:subspace_embedding}
For a matrix $A \in \R^{n \times d}$, we say matrix $S$ is a Subspace Embedding $\mathsf{SE}(n,d,\epsilon,\delta)$ of $A$ if
\begin{align*}
    \Pr[ \| S A x \|_2^2 = (1\pm \epsilon) \| A x\|_2^2 , \forall x \in \R^d] \ge  1-\delta.
\end{align*} 
\end{definition}

We define approximate matrix product (see \cite{w14} for example) as follows.
\begin{definition}[Frobenius Approximate Matrix Product]\label{def:approximate_matrix_product}
  Let $\epsilon \in (0,1)$ denote an accuracy parameter. Let $\delta \in (0,1)$ denote a failure probability. Given matrices $A$ and $B$, where $A$ and $B$ each have $n$ rows. We say a matrix $S$ is a Frobenius Norm Approximate Matrix Product ($\mathsf{FAMP}(n, \epsilon, \delta)$) with respect to $A$, $B$ if 
  \begin{align*}
      \Pr[\|A^\top B - A^\top S^\top S B\|_F \le \epsilon \cdot \|A\|_F \cdot \|B\|_B] \ge 1 - \delta. 
  \end{align*}

\end{definition}
\subsection{Gaussian Transforms and CountSketch}
\label{sec:gauss_countsketch}
We define the following kinds of sketching matrices. 
\begin{definition}[$\mathsf{CountSketch}$ or Sparse Embedding matrix]
\label{def:count_sketch_transform}
    We define the following,
    \begin{itemize}
        \item $\sigma \in \R$: a scalar;
        \item $D \in \R^{n \times n}$: a diagonal matrix, with each diagonal entry chosen independently from $\{-1,+1\}$ with equal probability; 
        \item $h:[n] \rightarrow [m]$: a random map, where $\forall i \in [n], j \in [m]$, we define
        \begin{align*}
            \Pr[h(i) = j] = 1/m.
        \end{align*}
        \item $\Pi \in \{0, 1\}^{m \times n}$: a binary matrix, where for all $(j, i) \in [m] \times [n]$, the entries are defined as
        \begin{align*}
            \Phi_{j,i} := 
            \left\{
            \begin{array}{lc}
                 1&  \mathrm{if~}h(i) = j\\
                 0&  \mathrm{otherwise}
            \end{array}
            \right.
        \end{align*}
    \end{itemize}
    Then, the CountSketch matrix $\Pi \in \R^{m \times n}$ is defined as
    \begin{align*}
        \Pi := \sigma \cdot \Phi D.
    \end{align*}
    For any real matrix $A \in \R^{n \times d}$, the matrix product $\Pi A$ can be computed in input-sparsity $O(\nnz(A))$ time. For any real tensor $A \in \R^{n \times d_1 \times d_2}$, the product $\Pi A$ can be computed in input-sparsity $O(
    \nnz(A))$ time. For three CountSketch matrices $\Pi_i, \Pi_2, \Pi_3$, for any input $3$-order tensor $A \in \R^{n_1 \times n_2 \times n_3}$, $A(\Pi_1, \Pi_2, \Pi_3)$ can be computed in input-sparsity $O(\nnz(A))$ time.

\end{definition}
In the absence of specification, the scalar $\sigma$ is assumed to be $1$ in the above notation.

\begin{definition}[$\mathsf{Gaussian}$ transform or $\mathsf{Gaussian}$ matrix ]
\label{def:gaussian_transform}
    Let $\sigma \in \R$ be a scalar. Let $G \in \R^{m \times n}$ be a matrix such that each entry of it is chosen i.i.d. from $\N(0, 1)$. Then we define $S = \sigma \cdot G \in \R^{m \times n}$ to be the Gaussian matrix or the Gaussian transform. For any input matrix $A \in \R^{n \times d}$, the matrix product $SA$ can be computed in input sparsity time of $O(m \cdot \nnz(A))$ time. For any input tensor, the product can be computed in input-sparsity $O(m \cdot \nnz(A))$ time.

\end{definition}
In the absence of specification, the scalar $\sigma$ is assumed to be $1/\sqrt{m}$ in the above notation.

We use the $\mathsf{CountSketch}$ together Gaussian transforms to get the following transform:

\begin{definition}[$\mathsf{CountSketch}$ + $\mathsf{Gaussian}$ transform]\label{def:fast_gaussian_transform}
    Let $\Pi \in \R^{t\times n}$ denote a $\mathsf{CountSketch}$ matrix (Definition~\ref{def:count_sketch_transform}). Let $S \in \R^{m \times t}$ denote a Gaussian matrix (Definition~\ref{def:gaussian_transform}). We define $S' = S \Pi$. Then we have that, for any input matrix $A \in \R^{n \times d}$, the matrix product $S'A$ can be computed in input-sparsity $O(\nnz(A) + dtm^{\omega-2})$ time, where we use $\omega$ to denote the matrix multiplication exponent.
    
\end{definition}
For the above sketching matrix, we have the following useful result telling us its $\mathsf{SE}$ and $\mathsf{FAMP}$ properties.  
\begin{lemma}[Lemma B.22 in \cite{swz19}
]\label{lem:gaussian_sketch_for_regression}\label{lem:gaussian_count_sketch_for_regression}
    Let $m_2=\Omega(k^2+k/\varepsilon)$, $m_1=\Omega(k/\varepsilon)$. Let $\Pi\in\R^{m_2\times n}$ be a CountSketch matrix (Definition~\ref{def:count_sketch_transform}). Let $S \in \R^{m_1 \times m_2}$ be a Gaussian matrix (Definition~\ref{def:gaussian_transform}). Then we define $S'$ using $S\Pi$ as Definition~\ref{def:fast_gaussian_transform}. Then  for any fixed matrix $C\in\R^{n\times k}$, $S'$ is an $\mathsf{SE}(n,k,1/3,0.01)$ (Definition~\ref{def:subspace_embedding}). Also, for any two fixed matrix $A$ and $B$ with the same number of rows, it is an $\mathsf{FAMP}(n, O(\epsilon/k), 0.01)$ (Definition~\ref{def:approximate_matrix_product}) .

\end{lemma}

\begin{theorem}[Theorem 36 in~\cite{cw13}]\label{thm:multiple_regression_sketch}
    Given $A \in \R^{n \times k},B \in \R^{n \times d},$ suppose $S\in\R^{m\times n}$ is an $\mathsf{SE}(n,k,1/\sqrt{2},\delta)$ for $A$, and satisfies is an $\mathsf{FAMP}(n, O(\sqrt{\epsilon/k},\delta))$ matrices $A$ and $C$ where $C$ has $n$ rows and $C$ depends on $A$ and $B$. If
    \begin{align*}
        \wh{X}=\arg\min_{X \in \R^{k\times d} } \|SB-SAX\|_F^2,
    \end{align*}
    then
    \begin{align*}
        \|B-A\wh{X}\|_F^2\leq (1+\varepsilon)\min_{X\in\R^{k\times d}}\|B-AX\|_F^2.
    \end{align*}
\end{theorem}

\subsection{Tensor Low-rank Approximation}
\label{sec:f_1_plus_epsilon}

In this section, we provide the following algorithm for tenor low-rank approximation. 

\begin{algorithm}[ht]\caption{An algorithm for solving Frobenius norm low-rank approximation problem}\label{alg:f_main_algorithm}
\begin{algorithmic}[1]
\Procedure{\textsc{FLowRankApprox}}{$A,n,k,\epsilon$} \Comment{Theorem \ref{thm:f_main_algorithm}}

\For{$i=1 \to 3$}
    \State $s_i \gets O(k/\epsilon)$.
\EndFor
\State Select sketching matrices $S_1\in \R^{n^2 \times s_1}$, $S_2 \in \R^{n^2 \times s_2}$, $S_3\in \R^{n^2 \times s_3}$. \Comment{Definition \ref{def:fast_gaussian_transform}}
\For{$i=1 \to 3$}
\State Compute $A_i S_i$.  \label{sta:f_main_compute_AiSi}
\EndFor
\State $Y_1, Y_2, Y_3, C \gets$\textsc{FInputSparsityReduction}($A,A_1S_1,A_2S_2,A_3S_3,n,s_1,s_2,s_3,k,\epsilon$). \label{sta:f_main_input_sparsity_reduction} \Comment{Algorithm~\ref{alg:f_input_sparsity_reduction}}
\For{$i=1 \to 3$}
    \State Create variables for $X_i \in \R^{s_i \times k}$
\EndFor
\State Run polynomial system verifier for $\| (Y_1 X_1) \otimes (Y_2 X_2) \otimes (Y_3 X_3) -C\|_F^2$. \label{sta:f_main_solve_small_problem}
\State \Return $A_1 S_1 X_1$, $A_2 S_2 X_2$, and $A_3S_3 X_3$.
\EndProcedure
\end{algorithmic}
\end{algorithm}

\begin{theorem}[Theorem~C.1 in page 31 in  \cite{swz19}]\label{thm:f_main_algorithm}
If the following conditions holds
\begin{itemize}
\item Let $n$ denote a positive integer.
\item Let $A\in \R^{n\times n \times n}$ be a 3rd order tensor.
\item For any integer $k\geq 1$.
\item For any accuracy parameter $\epsilon\in(0,1)$.
\end{itemize}
Then, an algorithm exists such that
\begin{itemize}
\item 
runs in $\exp( O(k^2/\epsilon)+n \poly(k,1/\epsilon) +O(\nnz(A))$ time 
\item construct three matrices $U$, $V$, and $W$ of dimension $n \times k$.
such that
\begin{align*}
\left\| \sum_{i=1}^k U_i \otimes V_i \otimes W_i -A \right\|_F^2 \leq (1+\epsilon) \underset{\rank-k~A_k}{\min} \| A_k -A \|_F^2
\end{align*}
\item the succeed probability is $9/10$.
\end{itemize}
\end{theorem}

\subsection{Input Sparsity Reduction}
\label{sec:f_input_sparsity_reduction}
Here in this section, we introduce following key lemma for achieving input-sparsity running time. 
\begin{algorithm}[ht]\caption{In this algorithm, we reduce the size of the  objective function's from $\poly(n)$ size to $\poly(k)$ size}\label{alg:f_input_sparsity_reduction}
\begin{algorithmic}[1]
\Procedure{\textsc{FInputSparsityReduction}}{$A,V_1,V_2,V_3,n,b_1,b_2,b_3,k,\epsilon$} \Comment{Lemma \ref{lem:f_input_sparsity_reduction}}
\For{$i=1 \to 3$}
    \State $c_i \gets \poly(k,1/\epsilon)$.
\EndFor
\For{$i= 1 \to 3$}
\State Choose sparse embedding matrices $T_i\in \R^{c_i \times n}$
\Comment{Definition~\ref{def:count_sketch_transform}}
\EndFor 
\For{$i=1 \to 3$}
    \State $\wh{V}_i \gets T_i V_i \in \R^{c_i \times b_i}, $ 
\EndFor
\State $C\gets A(T_1,T_2,T_3) \in \R^{c_1\times c_2 \times c_3}$.
\State \Return $\wh{V}_1$, $\wh{V}_2$, $\wh{V}_3$ and $C$.
\EndProcedure
\end{algorithmic}
\end{algorithm}

\begin{lemma}[Lemma~C.3 in page 35 in \cite{swz19}]\label{lem:f_input_sparsity_reduction}
If the following condition holds
\begin{itemize}
\item  Let $\poly(k,1/\epsilon) \geq b_1b_2b_3\geq k$. 
\item Suppose there are three matrices $V_1\in \R^{n\times b_1}$, $V_2 \in \R^{n\times b_2}$, and $V_3 \in \R^{n\times b_3}$.
\item Given a $n \times n \times n$ size tensor $A $
\item Let ${\cal T} = O(\nnz(V_1) + \nnz(V_2) + \nnz(V_3)+\nnz(A))=O(n\poly(k/\varepsilon)+\nnz(A))$
\item Let $c_1=c_2=c_3=\poly(k,1/\epsilon)$
\end{itemize}
Suppose there is an algorithm that
\begin{itemize}
\item runs in ${\cal T}$ time 
\item Constructs a tensor $C\in \R^{c_1\times c_2\times c_3}$ 
\item Constructs three matrices $\wh{V}_1\in \R^{c_1\times b_1}$, $\wh{V}_2 \in \R^{c_2\times b_2}$ and $\wh{V}_3 \in \R^{c_3 \times b_3}$ 
\item with probability at least $0.99$, for all $\alpha>0,X_1,X'_1\in\R^{b_1\times k}, X_2,X'_2\in\R^{b_2\times k}, X_3,X'_3\in\R^{b_3\times k}$ satisfy that,{\small
\begin{align*}
\left\| \sum_{i=1}^k (\wh{V}_1 X_1')_i \otimes (\wh{V}_2 X_2')_i \otimes (\wh{V}_3 X_3')_i - C \right\|_F^2 \leq \alpha \left\| \sum_{i=1}^k (\wh{V}_1 X_1)_i \otimes (\wh{V}_2 X_2)_i \otimes (\wh{V}_3 X_3)_i - C \right\|_F^2,
\end{align*}}
\end{itemize}
Then,{\small
\begin{align*}
\left\| \sum_{i=1}^k (V_1 X_1')_i \otimes (V_2 X_2')_i \otimes ( V_3 X_3')_i - A \right\|_F^2 \leq (1+\epsilon) \alpha \left\| \sum_{i=1}^k ({V}_1 X_1)_i \otimes ({V}_2 X_2)_i \otimes ({V}_3 X_3)_i - A \right\|_F^2.
\end{align*}}

\end{lemma}

\subsection{Solving Small Problems}
\label{sec:f_solving_small_problems}
Here we introduce the following theorem for solving small problem of our algorithm.
\begin{theorem}[Theorem~D.11 in page 90 in  \cite{swz19}]\label{thm:f_solving_small_problems}
If the following conditions hold
\begin{itemize}
    \item Suppose we have $\max \{t_1, d_1, t_2, d_2, t_3, d_3\} \leq n$.
    \item Let $A$ be a tensor that has size $t_1 \times t_2 \times t_3$
    \item Given three matrices
    \begin{itemize}
        \item $t_1 \times d_1$ matrix $T_1$
        \item $t_2 \times d_2$ matrix $T_2$
        \item $t_3 \times d_3$ matrix $T_3$
    \end{itemize} 
    \item for any $\delta > 0$ there exists a solution to
\begin{align*}
\min_{X_1,X_2,X_3} \left\| \sum_{i=1}^k (T_1 X_1)_i \otimes (T_2 X_2)_i \otimes (T_3 X_3)_i - A \right\|_F^2 := \OPT,
\end{align*}
and representing each element of $X_i$ using $O(n^\delta)$ bits is achievable.
\end{itemize}
Then there is an algorithm that
\begin{itemize}
\item runs in $ n^{O(\delta)} \cdot \exp( O( d_1 k+d_2 k+d_3 k))$ time
\item constructs three matrices
\begin{itemize}
    \item $\wh{X}_1$ 
    \item $\wh{X}_2$ 
    \item $\wh{X}_3$ 
\end{itemize}
\item such that $\| (T_1 \wh{X}_1)\otimes (T_2 \wh{X}_2) \otimes (T_3\wh{X}_3) - A\|_F^2 =\OPT$.
\end{itemize}
\end{theorem}

\section{Tensor Cycle Rank}
\label{sec:tensor_rank}
In this section, we provide our main algorithm and its analysis. 

\begin{algorithm}[ht]\caption{Our Main Algorithm
}\label{alg:main}
\begin{algorithmic}[1]
\Procedure{\textsc{FLowCycleRankApprox}}{$A,n,k,\epsilon$} \Comment{Theorem \ref{thm:main_formal}}
\For{$i=1 \to 3$}
\State $s_i  \gets O(k^2/\epsilon)$. 
\State $t_i \gets \poly(k,1/\epsilon)$.
\State  Choose sketching matrices $S_i\in \R^{n^2 \times s_i}$ \Comment{Definition \ref{def:fast_gaussian_transform}}
\State  Choose sketching matrices $T_i\in \R^{t_i \times n}$
\EndFor

\For{$i=1\to3$}\label{sta:train_compute_AiSi}
\State Compute $A_i S_i$
\EndFor
\For{$i=1\to 3$}\label{sta:train_compute_TiAiSi}
\State Compute $T_i A_i S_i$
\EndFor
\State Compute $B \gets A(T_1,T_2,T_3)$. \label{sta:train_compute_AT1T2T3}
\For{$i=1 \to 3$}
    \State Create variables for $X_i \in \R^{s_i \times k^2}$.
\EndFor

\State Create variables for $C \in \R^{k^2 \times k^2 \times k^2}$.  
\State $D\gets \|\sum_{i_1=1}^k \sum_{i_2=1}^k \sum_{i_3=1}^k (Y_1 X_1)_{i_1+k(i_2-1)} (Y_2 X_2)_{i_2 + k(i_3-1)}  (Y_3 X_3)_{i_3+k(i_1-1)}  - B\|_F^2$ \label{line:solve}
\State Run polynomial system verifier for D.\label{sta:train_solve_small_problem}
\State \Return $A_1 S_1 X_1$, $A_2 S_2 X_2$, and $A_3S_3 X_3$.
\EndProcedure
\end{algorithmic}
\end{algorithm}

\begin{theorem}[Cycle rank, formal version of Theorem~\ref{thm:main_informal}]\label{thm:main_formal}
For any $k\geq 1$, $\epsilon \in (0,1)$, and a third-order tensor $A\in \R^{n\times n\times n}$, there is an algorithm that operates in
\begin{align*}
    O(\nnz(A)) + n \poly(k,1/\epsilon) + \exp( O(k^6/\epsilon) )
\end{align*}
time and produces three tensors $U$, $V$, $W$ of dimension $n \times k \times k$ satisfying the following condition:
\begin{align*}
\| \sum_{i=1}^k \sum_{j=1}^k \sum_{l=1}^k U_{*,i,j} \otimes V_{*,j,l} \otimes W_{*,l,i} - A  \|_F \leq (1+\epsilon) \underset{\mathrm{cycle}~\rank-k~ A_k }{\min} \| A_k - A\|_F^2
\end{align*}
or equivalently
\begin{align*}
\left\| \sum_{i=1}^k \sum_{j=1}^k \sum_{l=1}^k (U_2)_{*,i + k (j-1)} \otimes (V_2)_{*, j + k(l -1) } \otimes (W_2)_{*, l+k(i-1)} - A \right\|_F^2 \leq (1+\epsilon) \underset{\mathrm{cycle}~\rank-k~ A_k }{\min} \| A_k - A\|_F^2
\end{align*}
holds with probability $9/10$. 

Here 
\begin{itemize}
\item 
$U_2 \in \R^{n \times k^2}$ is the flatten version of tensor $U \in \R^{n \times k \times k}$.
\item $V_2 \in \R^{n \times k^2}$ is the flatten version of tensor $V \in \R^{n \times k \times k}$.
\item $W_2 \in \R^{n \times k^2}$ is the flatten version of tensor $W \in \R^{n \times k \times k}$.
\end{itemize}
\end{theorem}

\begin{proof}

We define $\OPT$ as
\begin{align*}
\OPT: =\min_{\mathrm{cycle}~\rank-k~A'} \| A' -A \|_F^2.
\end{align*}

Suppose the optimal
\begin{align*}
A_k=  \sum_{i=1}^k \sum_{j=1}^k \sum_{l=1}^k U^*_{i+k(j-1)} \otimes V^*_{ j + k(l-1)} \otimes W^*_{l+k(i-1)}.
\end{align*}

Let $V^* \in \R^{n \times k^2}$ and $W^* \in \R^{n \times k^2}$ be fixed. We refer to the columns of $V^*$ as $V_1^*, V_2^*, \cdots, V_{k^2}^* \in \R^n$ and the columns of $W^*$ as $W_1^*, W_2^*, \cdots, W_{k^2}^* \in \R^n$.

The optimization problem under consideration is as follows:
\begin{align*}
\min_{U \in \R^{n\times k^2} } \left\|  \sum_{i=1}^k \sum_{j=1}^k \sum_{l=1}^k  U_{i+k(j-1)} \otimes V^*_{ j + k(l-1)} \otimes W^*_{l + k (i-1)} - A \right\|_F^2,
\end{align*}
which can be rewritten as
\begin{align*}
\min_{U \in \R^{n \times k^2} } \left\| U \cdot
\begin{bmatrix}
\sum_{l=1}^k V_{1 + k(l-1)}^* \otimes W^*_{l + k \cdot 0} \\
\sum_{l=1}^k V_{2 + k(l-1)}^* \otimes W^*_{l + k \cdot 0} \\
\vdots \\
\sum_{l=1}^k V_{k + k(l-1)}^* \otimes W^*_{l + k \cdot 0} \\
\sum_{l=1}^k V_{1 + k(l-1)}^* \otimes W^*_{l + k \cdot 1} \\
\sum_{l=1}^k V_{2 + k(l-1)}^* \otimes W^*_{l + k \cdot 1} \\
\vdots \\
\sum_{l=1}^k V_{k + k(l-1)}^* \otimes W^*_{l + k \cdot 1} \\
\sum_{l=1}^k V_{1 + k(l-1)}^* \otimes W^*_{l + k \cdot (k-1)} \\
\sum_{l=1}^k V_{2 + k(l-1)}^* \otimes W^*_{l + k \cdot (k-1)} \\
\vdots \\
\sum_{l=1}^k V_{k + k(l-1)}^* \otimes W^*_{l + k \cdot (k-1)}
\end{bmatrix} -A \right\|_F^2.
\end{align*}
We can represent a tensor $A \in \R^{n \times n \times n}$ as a matrix $A_1 \in \R^{n\times n^2}$ by flattening the tensor along the first dimension.
Matrix $Z_1 \in \R^{k^2 \times n^2}$ is defined as follows:
\begin{align*}
\begin{bmatrix}
\vect( \sum_{l=1}^k V_{1 + k(l-1)}^* \otimes W^*_{l + k \cdot 0} ) \\
\vect( \sum_{l=1}^k V_{2 + k(l-1)}^* \otimes W^*_{l + k \cdot 0} ) \\
\vdots \\
\vect( \sum_{l=1}^k V_{k + k(l-1)}^* \otimes W^*_{l + k \cdot 0} ) \\
\vect( \sum_{l=1}^k V_{1 + k(l-1)}^* \otimes W^*_{l + k \cdot 1} ) \\
\vect( \sum_{l=1}^k V_{2 + k(l-1)}^* \otimes W^*_{l + k \cdot 1} ) \\
\vdots \\
\vect( \sum_{l=1}^k V_{k + k(l-1)}^* \otimes W^*_{l + k \cdot 1} )\\
\vect( \sum_{l=1}^k V_{1 + k(l-1)}^* \otimes W^*_{l + k \cdot (k-1)} )\\
\vect( \sum_{l=1}^k V_{2 + k(l-1)}^* \otimes W^*_{l + k \cdot (k-1)} )\\
\vdots \\
\vect( \sum_{l=1}^k V_{k + k(l-1)}^* \otimes W^*_{l + k \cdot (k-1)} )
\end{bmatrix}
\end{align*}
Now we get the equivalent objective function as follows, 
\begin{align*}
\min_{U \in \R^{n\times k^2} } \| U Z_1  - A_1 \|_F^2.
\end{align*}
With  $A_k=U^*Z_1$, we can have $\min_{U \in \R^{n\times k^2} } \| U Z_1  - A_1 \|_F^2=\OPT$.

We can formulate the following optimization problem by using a sketching matrix $S_1^\top \in \R^{s_1 \times n^2}$ defined in Definition~\ref{def:fast_gaussian_transform}, where $s_1 = O(k^2/\epsilon)$:
\begin{align*}
\min_{U \in \R^{n\times k^2} } \| U Z_1 S_1 - A_1 S_1 \|_F^2.
\end{align*}
The optimal solution to the optimization problem above is denoted by $\wh{U} \in \R^{n\times k^2}$.

Then, we have
\begin{align*}
\wh{U} = A_1 S_1 (Z_1 S_1)^\dagger.
\end{align*}

By Lemma~\ref{lem:gaussian_count_sketch_for_regression} and Theorem~\ref{thm:multiple_regression_sketch}, we have

\begin{align*}
\| \wh{U} Z_1  - A_1  \|_F^2 
\leq & ~ (1+\epsilon) \cdot \min_{U\in \R^{n\times k^2}} \| U Z_1 - A_1 \|_F^2 \\
= & ~ (1+\epsilon) \cdot \OPT,
\end{align*}

which means that
\begin{align*}
\left\| \sum_{i=1}^k \sum_{j=1}^k \sum_{l=1}^k \wh{U}_{i+ k(j-1) } \otimes V^*_{j+ k(l-1)} \otimes W^*_{l+k(i-1)} - A \right\|_F^2 \leq (1+\epsilon) \cdot \OPT.
\end{align*}
In order to write down $\wh{U}_1, \cdots, \wh{U}_{k^2} \in \R^n$, considering the matrix $A_1$ is given, and we create $s_1 \times k^2$ variables for matrix $(Z_1 S_1)^\dagger$.

Now in the second step, let the two matrices $\wh{U} \in \R^{n\times k^2}$ and $W^* \in \R^{n\times k^2}$ to be fixed, we convert the tensor $A$ into the matrix $A_2$. 

We use $Z_2 \in \R^{k^2\times n^2}$ to denote the matrix where for each $j \in [k]$ and $l \in [k]$, the $(j,l)$-th row is  
\begin{align*} 
    \vect(\sum_{i=1}^k \wh{U}_{i+k(j-1)} \otimes W^*_{l+k(i-1)}).
\end{align*}

Now we focus on the objective function: 
\begin{align*}
\min_{ V \in \R^{n\times k^2} } \| V Z_2 -A_2  \|_F^2,
\end{align*}
where $(1+\epsilon) \cdot \OPT$ is the upper bound of its optimal cost.

Given that $s_2=O(k^2/\epsilon)$, we use $S_2^\top \in\R^{s_2\times n^2}$ as a sketching matrix defined in Definition~\ref{def:fast_gaussian_transform}. We apply $S_2$ to the right hand side of the objective function and we get the following new objective function,
\begin{align*}
\min_{ V \in \R^{n\times k^2} }  \| V Z_2 S_2 - A_2 S_2 \|_F^2.
\end{align*}

We use $\wh{V} \in \R^{n\times k^2}$ as the optimal solution above. 

Now, we have
\begin{align*} 
\wh{V} = \underbrace{ A_2 }_{n \times n^2} \underbrace{ S_2 }_{ n^2 \times s_2 } \underbrace{ (Z_2 S_2)^\dagger }_{s_2 \times k^2}.
\end{align*}

By Lemma~\ref{lem:gaussian_count_sketch_for_regression} and Theorem~\ref{thm:multiple_regression_sketch}, we have,
\begin{align*}
\| \wh{V} Z_2 - A_2 \|_F^2 
\leq & ~ (1+\epsilon ) \cdot \underset{V\in \R^{n\times k} }{\min} \| V Z_2  - A_2 \|_F^2 \\
\leq & ~ (1+\epsilon)^2 \cdot \OPT,
\end{align*}
which means that
\begin{align*}
\left\| \sum_{i=1}^k \sum_{j=1}^k \sum_{l=1}^k \wh{U}_{i + k(j-1) } \otimes \wh{V}_{j+k(l-1)} \otimes W^*_{l + k(i-1)} - A \right\|_F^2 \leq (1+\epsilon )^2 \OPT.
\end{align*}
In order to obtain the columns $\wh{V}_1, \cdots, \wh{V}_{k^2} \in \R^n$, recall the matrix $A_2 \in \R^{n^2 \times n}$ is given, we create $s_2\times k^2$ variables for matrix $(Z_2 S_2)^\dagger$. 

We set the matrices $\wh{U} \in \R^{n\times k^2}$ and $\wh{V}\in \R^{n \times k^2}$ to fixed values in the third step. The tensor $A\in \R^{n\times n \times n}$ is converted into matrix $A_3 \in \R^{n^2 \times n}$. Let matrix $Z_3\in \R^{k^2 \times n^2}$ denote
\begin{align*}
\begin{bmatrix}
\sum_{j=1}^k \vect(\wh{U}_{1 + k (j-1)} \otimes \wh{V}_{j + k \cdot 0}) \\
\sum_{j=1}^k \vect(\wh{U}_{2 + k (j-1)} \otimes \wh{V}_{j + k \cdot 0}) \\
\cdots \\
\sum_{j=1}^k \vect(\wh{U}_{k + k (j-1)} \otimes \wh{V}_{j + k \cdot 0}) \\
\sum_{j=1}^k \vect(\wh{U}_{1 + k (j-1)} \otimes \wh{V}_{j + k \cdot 1}) \\
\sum_{j=1}^k \vect(\wh{U}_{2 + k (j-1)} \otimes \wh{V}_{j + k \cdot 1}) \\
\cdots \\
\sum_{j=1}^k \vect(\wh{U}_{k + k (j-1)} \otimes \wh{V}_{j + k \cdot 1}) \\
\sum_{j=1}^k \vect(\wh{U}_{1 + k (j-1)} \otimes \wh{V}_{j + k \cdot (k-1)}) \\
\sum_{j=1}^k \vect(\wh{U}_{2 + k (j-1)} \otimes \wh{V}_{j + k \cdot (k-1)}) \\
\cdots \\
\sum_{j=1}^k \vect(\wh{U}_{k + k (j-1)} \otimes \wh{V}_{j + k \cdot (k-1)}) \\
\end{bmatrix}
.
\end{align*}
The objective function as follows is considered by us,
\begin{align*}
\underset{W\in \R^{n\times k^2} }{\min} \| W Z_3 - A_3 \|_F^2,
\end{align*}
with at most $(1+\epsilon)^2 \OPT$ optimal cost.

We can use a sketching matrix $S_3^\top \in\R^{s_3\times n^2}$, defined in Definition~\ref{def:fast_gaussian_transform}, with sketch size $s_3=O(k^2/\varepsilon)$, to reduce the size of the input. By sketching $S_3$ on the right of the objective function, we have
\begin{align*}
\underset{ W \in \R^{n \times k^2} }{ \min } \| W Z_3 S_3 - A_3 S_3 \|_F^2.
\end{align*}
Given that $\wh{W} \in \R^{n\times k^2}$ is used as the optimal solution, we can obtain $\wh{W} = A_3 S_3 (Z_3 S_3)^\dagger$.

By Lemma~\ref{lem:gaussian_count_sketch_for_regression} and Theorem~\ref{thm:multiple_regression_sketch}, we have,
\begin{align*}
\| \wh{W} Z_3 - A_3 \|_F^2 
\leq & ~ (1+\epsilon) \underset{W\in \R^{n\times k^2} }{\min} \| W Z_3 - A_3 \|_F^2 \\
\leq & ~ (1+\epsilon)^3 \OPT.
\end{align*}

Thus, we have
\begin{align*}
& ~ \min_{X_1,X_2,X_3} \left\| \sum_{i=1}^k \sum_{j=1}^k \sum_{l=1}^k (A_1 S_1 X_1)_{i + k (j-1) } \otimes (A_2S_2 X_2)_{j+k(l-1)} \otimes ( A_3S_3 X_3 )_{l+ k (i-1)}- A \right\|_F^2 \\
\leq & ~ (1+\epsilon)^3 \OPT.
\end{align*}
Let $V_1=A_1S_1,V_2=A_2S_2,$ and $V_3=A_3S_3.$ We then apply Lemma \ref{lem:f_input_sparsity_reduction}, and we obtain $\wh{V}_1,\wh{V}_2,\wh{V}_3,B$. We utilize Theorem~\ref{thm:f_solving_small_problems} to solve the problem efficiently. By rescaling $\epsilon$ by a constant factor, we can ensure the correctness of the algorithm.

\paragraph{Time complexity.} 

The running time of the algorithm can be divided as follows,
\begin{itemize}
    \item By Definition~\ref{def:fast_gaussian_transform}, Line~\ref{sta:train_compute_AiSi} in Algorithm~\ref{alg:main} takes time of
    \begin{align*}
        O(\nnz(A)) + n\poly(k,1/\epsilon);
    \end{align*}
    \item By Lemma~\ref{lem:f_input_sparsity_reduction}, Line~\ref{sta:train_compute_TiAiSi} and Line~\ref{sta:train_compute_AT1T2T3} takes time of 
    \begin{align*}
        \nnz(A) + n\poly(k,1/\epsilon);
    \end{align*}
    \item By Theorem~\ref{thm:f_solving_small_problems}, Line~\ref{line:solve} takes time of 
    \begin{align*}
        \exp( O(k^6/\epsilon) ),
    \end{align*}
    ignoring the bit complexity for simplicity. 
\end{itemize}
Adding then together, the total running time is
\begin{align*}
    O(\nnz(A)) + n \poly(k,1/\epsilon) + \exp( O(k^6/\epsilon) ).
\end{align*}
Thus we complete the proof. 
\end{proof}

\ifdefined\isarxiv
\bibliographystyle{alpha}
\bibliography{ref}
\else
\bibliography{ref}
\bibliographystyle{alpha}

\fi

\newpage
\onecolumn
\appendix




\end{document}